%% file: main.tex
\documentclass{article}
\usepackage[utf8]{inputenc}
\usepackage{titling}

\usepackage{enumerate}
\usepackage{lineno}

\usepackage{amssymb,ragged2e,booktabs,amsmath,amsfonts,eurosym,geometry,ulem,graphicx,caption,color,setspace,sectsty,comment,footmisc,pdflscape,array,subcaption}

\usepackage[labelsep=none,labelformat=empty]{caption}

\usepackage{forest}

\usepackage{tikz}
\usetikzlibrary{calc}
\usetikzlibrary{bayesnet}

\usepackage[square,sort,comma,numbers]{natbib}
\usepackage{bibentry}

\usepackage{tabularx,dcolumn,array,tabularx,booktabs}
\usepackage{epigraph,ragged2e}
\usepackage{hyperref}

\usepackage{tcolorbox}

\usepackage{blkarray}

\newcolumntype{L}[1]{>{\raggedright\let\newline\\arraybackslash\hspace{0pt}}m{#1}}
\newcolumntype{C}[1]{>{\centering\let\newline\\arraybackslash\hspace{0pt}}m{#1}}
\newcolumntype{R}[1]{>{\raggedleft\let\newline\\arraybackslash\hspace{0pt}}m{#1}}

\renewenvironment{abstract}
 {\small
  \begin{center}
  \bfseries \abstractname\vspace{-.5em}\vspace{0pt}
  \end{center}
  \list{}{%
    \setlength{\leftmargin}{1.0cm}% <---------- CHANGE HERE
    \setlength{\rightmargin}{\leftmargin}%
  }%
  \item\relax}
 {\endlist}

\title{From Probability to Consilience: \\ How Explanatory Values Implement Bayesian Reasoning}
\author{Zachary Wojtowicz\thanks{Department of Social and Decision Sciences, Carnegie Mellon University, Pittsburgh, PA} \and Simon DeDeo\footnotemark[1]\hspace{0.2cm}\thanks{Santa Fe Institute, Santa Fe, NM}}
\date{}

\begin{document}
%\nobibliography*

\maketitle
%\linenumbers

\begin{abstract}
    Recent work in cognitive science has uncovered a diversity of \textit{explanatory values}, or dimensions along which we judge explanations as better or worse. We propose a Bayesian account of how these values fit together to guide explanation. The resulting taxonomy provides a set of predictors for which explanations people prefer and shows how core values from psychology, statistics, and the philosophy of science emerge from a common mathematical framework. In addition to operationalizing the explanatory \textit{virtues} associated with, for example, scientific argument-making, this framework also enables us to reinterpret the explanatory \textit{vices} that drive conspiracy theories, delusions, and extremist ideologies.
\end{abstract}
    % The resulting taxonomy provides a set of predictors for when people adopt explanations, and shows how core values from psychology, statistics, and the philosophy of science emerge from a common mathematical framework.

\input{highlights}
% Outstanding Questions Box (2000 characters, including spaces, required)
% ¥ Important questions for future research should be summarized in a box (not included in box count or element limit). This is an excellent opportunity to offer input and guidance on new directions for the field.
% ¥ Please write succinct questions in list format, with bullet points to indicate the start of a new concept. 
% ¥ The Outstanding Questions Box should not include references.
% ¥ The box should be called out in an appropriate section in the text, generally the Concluding Remarks section, as 'see Outstanding Questions'. This element will be placed as the last box in the paper, although it should not be numbered with the other boxes.
% ¥ When submitting the manuscript files via Editorial Manager, please upload the Outstanding Questions Box as a separate word file using the designated heading.

\section{Explaining explanation}

%% need to cut 168 words :( -- working on it
Intuitively, philosophically, and as seen in laboratory experiments, {\bf explanations} are judged as better or worse on the basis of many different criteria. These {\bf explanatory values} appear in early childhood \citep{lombrozo2016explanatory,walker2017effects,samarapungavan1992children,frazier2016young,bonawitz2012occam} and their influence extends to some of the most sophisticated social knowledge formation processes we know \cite{harman1965inference}. We lack, however, an understanding of the origin of these values or an account of how they fit together to guide belief formation. The multiplicity of values also appears to conflict with Bayesian models of cognition, which speak solely in terms of degrees of beliefs and suggest we judge explanations as better or worse on the basis of a single quantity, the posterior {\bf likelihood} (see Glossary). In this opinion, we show how to resolve these conflicts by arguing that previously-identified explanatory values capture different components of a full Bayesian calculation and, when considered together and weighed appropriately, implement Bayesian cognition. %...inference over explanations? 

This framework shows how key explanatory values identified by laboratory experiments and philosophers of science---{\bf co-explanation}, {\bf descriptiveness}, {\bf precision}, {\bf unification}, {\bf power}, and {\bf simplicity}---emerge naturally from the mathematical structure of probabilistic inference, thereby reconciling them with Bayesian models of cognition \citep{griffiths2010probabilistic, sanborn2016bayesian}. Second, it shows how these values combine to produce preferences for one explanation over another. Third, it emphasizes new conceptual distinctions, such as one between explanatory values that can be assessed before the arrival of data ({\bf theoretical values}) and those that can only be assessed after the arrival of data ({\bf empirical values}). Finally, it enables us to reinterpret work on the characteristic deviations from normative patterns of explanation that drive phenomena such as conspiracy theories, delusions, and extremist ideologies.

% okasha2000van removed because we had to satisfy number of references request, and referee wanted a citation
It also resolves a tension in the influential philosophical account of ``inference to the best explanation'' (IBE; \citep{harman1965inference,schupbach2016inference,lombrozo2016explanatory}) which says belief formation is, or should be, guided by explanatory considerations. While some hold that IBE is incompatible with Bayesian updating because explanatory considerations cannot be captured within a probabilistic framework \citep{douven2017inference,douven2017inference2}, others argue that the two are either compatible \citep{climenhaga2017inference,huemer2009parsimony,lipton2004inference,weisberg2009locating} or potentially even identical \citep{henderson2014bayesianism,van1989laws}. We adopt this latter perspective, and show how our framework provides a compelling---albeit preliminary---account of how such an ``emergent compatabilism'' \cite{henderson2014bayesianism} can be achieved. \footnote{A more nuanced difference between Bayesianism and many variants of IBE is that the former, on its own, provides a system for determining the relative strength of various theories while the latter actually provides grounds for accepting a single, ``best'' explanation \citep{psillos2004inference}. Clearly, however, one can combine Bayesian updating with any number of rules for acceptance, such as choosing the explanation that has the maximum a posteriori probability (MAP) of being correct.} 

     \begin{figure}
        \centering
        \begin{tcolorbox}
            \textbf{Box 1: Bayes' Rule Decomposes into Explanatory Values} \vspace{3mm} \newline
            Formally, an explanation $E$ is evaluated on the basis of its log-likelihood in the presence of evidence $x = \{x_1,\ldots,x_n\}$; using Bayes rule, this gives us
            \begin{align*}
                % uses vphantom hack to vertically align underbraces
                \log p(E|x) = \underbrace{\vphantom{\sum_0^0}\sum_{i=1}^n\log p(x_i|E)}_{\substack{\text{Descriptiveness}}} + \underbrace{\vphantom{\sum_0^0}\log\left(\frac{p(x|E)}{\prod_{i=1}^n p(x_i|E)}\right)}_{\substack{\text{Co-explanation}}} + \underbrace{\sum_i\vphantom{\sum_0^0}\log T_i(E)}_{\substack{\text{Theoretical Values}}} + \underbrace{\vphantom{\sum_0^0}\log \pi(E)}_{\substack{\text{Contextual Factors}}}
            \end{align*}
            The terms in this decomposition are: (i) descriptiveness, which measures the total extent to which the explanation predicts each fact in isolation from the others; (ii) co-explanation, which measures the extent to which the explanation links facts together; (iii) theoretical, or evidence-independent values; and (iv) context-dependent priors. (For simplicity, we do not show the additional normalization term that is constant across all explanations and thus does not affect comparisons between explanations.)
         \end{tcolorbox}
         \caption{}
         \label{box:decomposition}
     \end{figure} 
     
\section{A Bayesian Framework for Explanatory Values}

Bayes' rule says that we should value an explanation in terms of a ``posterior'' degree of belief, determined by our prior degree of belief in that explanation times the probability that it assigns to the data we have observed. Working with log-probabilities means that the components of this calculation combine additively, matching the intuition that we weigh multiple features of an explanation by adding them together to make final determinations about its validity.
 
\begin{figure}
    \centering
    \begin{tcolorbox}
    {\Large \textbf{Glossary}} \vspace{3mm} \\
    \textbf{Explanation}: an account of some observable aspect of the world. In the Bayesian framework, an explanation supplies a probability distribution over events. \vspace{2mm} \\
    \textbf{Explanatory Values}: explanatory features that lead us to prefer one explanation over another. \\
    \\
    \underline{\large \textbf{Empirical Values}}: \\
    (Ways in which an explanation can be valued on the basis of data.) \vspace{2mm}  \\
    \textbf{Log-Likelihood}: $\log p(x|E)$ \\
    The log-probability of observed data given an explanation. \vspace{3mm} \\
    \textbf{Descriptiveness}: $\sum_i \log p(x_i|E)$ \\
    The total log-probability of observed data given an explanation when each observation is considered in isolation. \vspace{3mm} \\
    \textbf{Co-explanation}: $\log\left(\frac{p(x|E)}{\prod_{i=1}^n p(x_i|E)}\right)$ \\
    The relative increase in log-probability that an explanation gives a pattern of observed data above its ability to predict each piece in isolation. Equal to point-wise mutual information in the case of two variables, or point-wise multi-information in the general case~\citep{studeny1998multiinformation}. \vspace{3mm} \\
    \\
    \underline{\large \textbf{Theoretical Values}}: \\
    (``Priors'', or ways in which an explanation can be valued without reference to data.) \vspace{2mm}  \\
    \textbf{Precision}: $\mathbb{E}_E[\log p(x|E)]$ \vspace{2mm} \\
    The expected likelihood of data conditional on the explanation being true. Also equal to the negative entropy of the explanation. Measures the degree to which an explanation's predictions concentrate in a particular subset of the space of all possible outcomes. \vspace{3mm} \\
    \textbf{Power}: $\mathbb{E}_E[\sum_i \log p(x_i|E)]$ \vspace{2mm} \\
    The expected descriptiveness of data conditional on the explanation being true. Measures the degree to which an explanation tends to produce individual pieces of data that it can account for in isolation. \vspace{3mm} \\
    \textbf{Unification}: $\mathbb{E}_E\left[\log\left(\frac{p(x|E)}{\prod_{i=1}^n p(x_i|E)}\right)\right]$ \vspace{2mm} \\
    The expected co-explanation of data conditional on the explanation being true. Also equal to the mutual information in the case of two variables, or the multi-information in the general case. Measures the degree to which an explanation predicts patterns of outcomes and connects multiple variables together. \vspace{3mm} \\
    \textbf{Simplicity}: any function that measures how straightforward an explanation is. Examples include parsimony, concision, and elegance. The appropriate choice of  will generally depend on context. \\
    \textbf{Parsimony}: a type of simplicity that reflects the number of elements, parameters, or principles an explanation requires. \\
    \textbf{Concision}: a type of simplicity that measures how compact an explanation is, \emph{e.g.}, by counting the number of words required to communicate it. \\
    \end{tcolorbox}
    \label{glossary}
 \end{figure}

Box~\ref{box:decomposition} shows how the log-posterior can be rearranged into a series of additive terms which define, mathematically, values identified across a variety of different contexts. The first two terms are empirical, and track how an explanation accounts for observed data: descriptiveness (formally, the log-likelihood under the assumption that the data are independent), measures how well an explanation predicts the facts in isolation, and co-explanation (formally, the information-theoretic pointwise multi-information), measures how well an explanation predicts patterns that connect facts together. The next terms are theoretical and track the value of an explanation independently from the data. As we will argue, two key theoretical values correspond to expected empirical values, while others reflect structural features of an explanation and context-depenent priors.

This leads to two pairs of explanatory values---descriptiveness and power, co-explanation and unification---that appear either at the empirical or the theoretical stage, respectively, along with an additional theoretical value of simplicity. The correspondences between the mathematical terms and the explanatory values are shown in the Glossary. We discuss each in turn.

\section{Explanation through the Lens of Descriptiveness and Power}

The simplest way to judge an explanation is to consider each piece of evidence for it independently, keeping a running tally of the degree to which it makes the explanation look better or worse. This is captured by descriptiveness, the sum of the independent log-probabilites of the relevant facts.\footnote{A similar approximation is frequently employed in statistics---the familiar identically and independently distributed (IID) assumption.}

Although descriptiveness neglects that facts are rarely independent, it nevertheless often works quite well. For example, when evaluating students on the basis of their grades, we can usually interpret each mark as an independent reflection of academic ability, thus making GPA a useful summary. On the other hand, overemphasis on descriptiveness in a domain where correlations really do matter results in a cognitive bias known as correlation neglect~\cite{enke2019correlation}.

The theoretical value corresponding to descriptiveness is power: how descriptive an explanation is in a world where it was true. Valuing power means valuing explanations that make more definite predictions. All other things being equal, more descriptiveness is always a good thing. Power is more ambiguous, and someone might consider power to be a virtue or a vice. High-power explanations make definitive predictions and therefore more easily falsified; they are also more easily learned from experience. Further, if you believe a high power explanation, you expect those who value descriptiveness to be receptive to it as well. 

Power can also be a vice, however, because the world is not always as predictable as we might wish. In uncertain situations, one might value low power explanations as more open-minded and allowing for a wider range of possibilities. Indeed, in statistics, Ref.~\cite{jaynes1979inference} has advocated for the ``principle of maximum entropy'', which views minimizing {\bf precision}---the sum of power and unification (discussed below)---as a universal normative rule of inference because it presumes to know the least \emph{a priori}.
% Is this, strictly speaking, true? now it is!

\begin{figure}
    \centering
    \begin{tcolorbox}
        \textbf{Box 2: Explanatory Values in Action} \vspace{3mm}
        
        \begin{center}
            \includegraphics[width=0.7\textwidth]{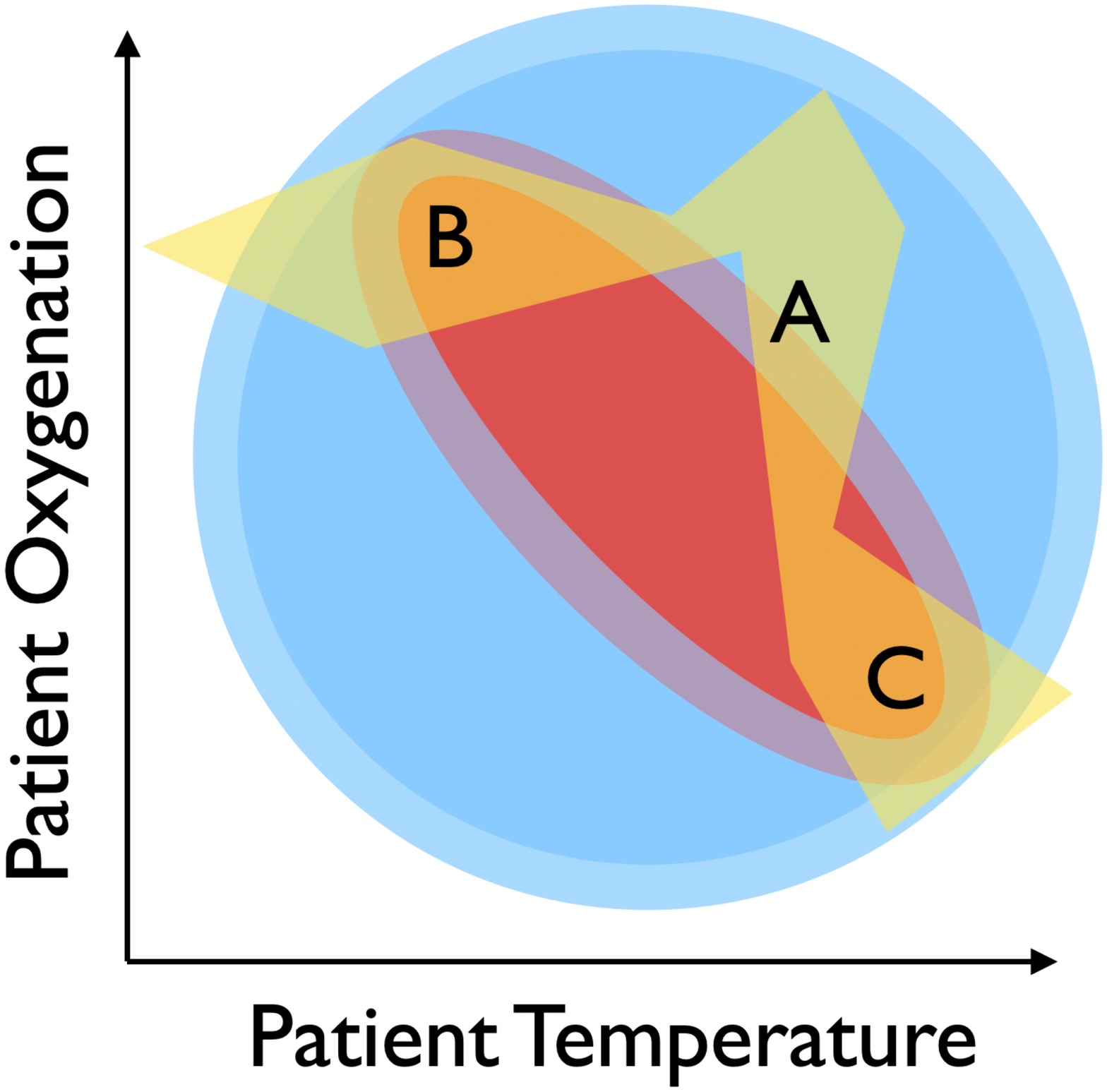}
        \end{center}
        
        A common paradigm to tease apart explanatory values is disease diagnosis~\citep{medin1982correlated,lombrozo2007simplicity,johnson2014explanatory}. Participants are asked to explain a patient's symptoms by reference to different medical conditions. The figure above illustrates a general case of this task, where there are three potential explanations (shaded in red, blue, and yellow, with density of each color indicating probability) that might produce different patterns of symptoms (here, different combinations of a patient's blood oxygenation and temperature). \\
        
        In the framework of Bayesian explanatory values proposed here, the blue explanation has low power (it allows for a wide range of outcomes) and low unification (patient temperature is not particularly well predicted by oxygenation). The red explanation has higher power (a narrower range of possibilities), and non-zero co-explanation (high patient temperatures are usually accompanied by low blood oxygenation). The yellow explanation is similar to the red in that it is both powerful and co-explanatory, but implies a less simple relationships between oxygenation and temperature. \\
        
        Confronted with three different patients (A, B, and C), these explanations also have different empirical values. For example, the red explanation has lower descriptiveness than the yellow explanation (patients A and C fall somewhat outside the normal range for red), but higher co-explanation than the blue (the temperatures of all three patients predict their oxygenation). For the particular case of patient A, red also has higher co-explanation than yellow, since a patient with A's temperature under the yellow explanation can have a wider oxygenation range. \\
                     
        Which explanation is best depends on context. Even if the yellow explanation is more valued on the basis of descriptiveness and co-explanation, power or unification, a person may still come to prefer the red---or even the blue---with a strong enough preference for simplicity. For example, the yellow explanation may produce its complex relationship between oxygenation and temperature by invoking the presence of two diseases simultaneously, or through a complicated interaction of different underlying conditions. \\
        
        Explanations are evaluated relative to a (usually stable) background ontology: here, oxygenation and temperature. If the ontology changes, so do the values and if, for example, doctors worked with a quantity equal to ``temperature minus oxygenation'', then the red explanation would become less co-explanatory and more descriptive, while their sum remains constant.
     \end{tcolorbox}
     \caption{}
     \label{box:example}
\end{figure} 
    
\section{Explanation through the Lens of Co-explanation and Unification}

In addition to considering facts in isolation, we also care about how they connect together. This is captured by ``co-explanation'', which measures how well an explanation predicts a pattern of observations over and above how well it accounts for each independently. While this definition arises naturally from our Bayesian decomposition, its mathematical form matches that proposed by Ref.~\citep{myrvold2003bayesian} as an operationalization of IBE and by Ref.~\citep{zhang2015likelihood} as an operationalization of explanatory considerations in Ref.~\citep{forster1988unification}. %, and is connected to some influential interpretations of Peirce's notion of abduction~\citep{}.

%%% Consilience of inductions???? is this supposed to be a ref. to Whewell? Check.

Co-explanation is high when an explanation says some features of the data are predicted by others. For example, when economists observed that unemployment and inflation were inversely correlated, they proposed that the relationship was a general law---the Phillips curve~\citep{phillips1958relation}. Explanations that included this law had, as a consequence, co-explanatory value: inflation appeared to become predictable given knowledge of unemployment. Beyond economics, this value is particularly relevant in domains characterized by correlating common causes, such as diseases in medical diagnosis \citep{dragulinescu2016inference}, legal cases \citep{amaya2016inference}, and social interactions~\citep{kleiman2016coordinate}; see Box 2.

Psychological studies in these domains are often implicit tests of an individual's sensitivity to co-explanation as an explanatory value. Ref.~\citep{medin1982correlated} trained participants on cases that noted the presence or absence of various symptoms for patients with a fictitious disease. When asked to judge which of two new patients was more likely to have the disease, subjects were sensitive to not only whether each of their symptoms was a likely result (descriptiveness), but also whether their presentation preserved correlations between symptoms seen in the training set (co-explanation).\footnote{Crucially, co-explanation requires variation along different dimensions: if it is impossible, for example, for the data vary along a particular axes under a certain explanation, than knowledge of that aspect of the world tells us nothing new about the others because there is nothing left to explain. In this special case, the theory may have many virtues, such as high descriptiveness or power, but its co-explanation is at a minimum.}
            
That the mind links distinct experiences into coherent wholes is also at the heart of Gestalt psychology: ``to apply the gestalt category means to find out which parts of nature belong as parts to functional wholes, to discover their position in these wholes, their degree of relative independence, and the articulation of larger wholes into sub-wholes''~\citep{koffka}. In vision, for example, it is argued that we perceive not the individual pieces of raw sense-data, but rather the explanation that links them together; the Kanizsa Triangle is compelling because the implicit shape we see co-explains the orientations of the missing wedges.
            
Co-explanation has the parallel theoretical value of unification, the expected co-explanation of the explanation conditional upon its truth. Unification says that the world is characterized not by coincidences, but patterns. It is a commonly value in the philosophy of science; for Ref.~\cite{friedman1974explanation}, a good scientific theory makes the manifestation of different phenomena dependent on each other, and a similar account can be found in Ref.~\cite{kitcher1989explanatory}, for whom good theories form a ``systemtic'' picture of the order of nature.

As with power, unification may be both virtue and vice. Even when a unifying theory is complicated, it does assert that the world itself is simple, because knowledge of some of its features allow you to predict others. Unifying theories, like powerful ones, are also more testable: one can look not only for unexpected events, but also  patterns.\footnote{More formally, the number of tests of a high power theory is linear in the number of features one looks at, but the number of tests of a unifying theory scale quadratically if one looks for pairwise correlations, or exponentially, when considering all combinations.} On the other hand, experiments show that unification may be perceived as a vice. Consider, for example, two explanations for ``why Lois painted her nails in the shower''~\cite{khemlani2011harry}: (a) `she is afraid of spilling nail polish on her antique bathroom rug'' or (b) ``she is obsessive--compulsive'',`. Explanation (b) correlates Lois' behavior with a many other (as yet) unobserved behaviors, and so is higher in unification (in that paper's terminology, has broader latent scope) than Explanation (a). However, subjects tended to prefer theories with lower unification (narrower latent scope).
% I think only worth including if we mention its findings and their significance
% \cite{johnson2014explanatory} expand on this work by introducing ``manifest'' as opposed to ``latent'' scope, roughly equivalent to the distinction between the values of co-explanation and unification.

\section{Explanation through the Lens of Simplicity and Other Priors}

Explanation and descriptiveness help us weigh explanations in a Bayesian fashion. Without simplicity, however, they are rarely good enough; as pointed out by thinkers such as Galileo, Newton, and Kant \cite{baker2004simplicity}, one can often ``improve'' an explanation by adding more parameters, exceptions, and mechanisms, an observation linked to Occam's Razor. In Bayesian inference, simplicity enters into the prior. Accordingly, our intuitive notion of simplicity is captured by one or more theoretical values \citep{good1968corroboration}.

Ref.~\cite{lombrozo2007simplicity} finds evidence for this using an alien disease paradigm that puts descriptiveness and simplicity (here, {\bf parsimony}) into conflict: participants' explanatory preferences are consistent with a value for descriptiveness plus a constant, \emph{i.e.} data-independent, penalty for theory elaborateness (lack of parsimony). Further research has shown that such preferences take form early in childhood development and are robust across contexts~\cite{bonawitz2012occam,walker2017effects,lombrozo2016explanatory}.

Simplicity is complex, and our intuitive notion is an amalgam of many theoretical considerations. Table~\ref{box:simplicity} illustrates the independent operation of two such considerations when explanations have a causal form. The upper-left theory is both more parsimonious (fewer hidden causes) and also more unified (providing a joint account of events) than the bottom-right theory. However, these effects can be decomposed. We therefore also have parsimonious-disjointed theories (where each visible aspect has a streamlined, but non-overlapping latent causal structure) and elaborate-unified theories (where everything is connected by a sprawling web of relations). This latter kind is reminiscent of conspiracy theories, where in one sense everything is very simple (all visible aspects are connected to each other), but that simplicity is achieved by postulating an elaborate web of hidden connections.

% Unification is not all, however: the first row of Box 3 might be said to be simpler (more parsimonious) than those in the second. The elaborate explanation of the lower-left theory are reminiscent of conspiracy theories, where in one sense everything is very simple (all visible aspects are connected to each other), but in another sense that simplicity (unification) is achieved by postulating numerous hidden connections.

% mackay1992bayesian reference cut to make <100
In statistics, simplicity is part of model selection, and there are many forms \citep{jaynes1979inference,mackay2003information}: the commonly-used Akaike Information Criterion \citep{akaike1974new} and Bayesian Information Criterion \citep{schwarz1978estimating} are parsimony measures that count the number of parameters in a theory. In machine learning, regularization terms penalize non-zero parameter values to prevent overfitting to data. Meanwhile, the maximum entropy principle \citep{jaynes2003probability} understands simplicity as (negative) precision. There is general consensus both (1) simplicity is crucial to normative decision making, and (2) however the value is measured, it ought to enter additively in log-space, \emph{i.e.}, as term in the prior shown in our Box~\ref{box:decomposition} \citep{good1977explicativity}.

While statistical models are often concerned with prediction, the psychological value of simplicity goes well beyond this goal. Simple explanations can be easier to remember and work with, and may be preferred because we are limited beings with cognitive constraints. A simple explanation may be better because it requires fewer cognitive resources to apply or leads to fewer mistakes. Simple explanations are also easier to communicate and teach, and thus ease social coordination. Ssimplicity may even be seen as an aesthetic value, with simple explanations described as ``elegant'' and, on that basis, valuable in and of themselves; ``mathematical beauty'' plays a significant~\cite{greene2000elegant,scarry2013beauty}, if controversial~\cite{hossenfelder2018lost}, role in physics.

Simplicity is not the only domain-general theoretical value, as there are many prior reasons to prefer one theory over another. Entire classes of explanations may be categorically better than others; for example, those that reveal causal mechanisms~\cite{salmon1998causality,zemla2017evaluating,zemla2020not}, explain new phenomena by analogy to familiar ones \citep{thagard1978best}, or feature concrete mechanisms rather than abstract principles \cite{dragulinescu2016inference}. Within each class, humans have been shown to hold strong, contextually-informed priors for certain kinds of explanations. For example, we prefer explanations that involve diseases causing symptoms over the reverse \citep{tenenbaum2007intuitive} and find some causal connections immediately implausible \citep{yeung2015identifying}. These priors reflect our ability to flexibly apply background knowledge to new problems \citep{norton2003material} and often take the form of intuitions and instincts---what Galileo referred to as ``il lume naurale'' or ``the natural light'' that guides our reason (1.80 of Ref.~\cite{peirce1960collected}).

\begin{figure}
    \centering
    % \begin{tcolorbox}
        % \textbf{Box 2: Different Axes of Simplicity} \newline
        \begin{center}
        \begin{tabular}{ r|>{\centering\arraybackslash} m{5.5cm}| >{\centering\arraybackslash} m{5.5cm}| }
        \multicolumn{1}{c}{} & \multicolumn{1}{c}{\textbf{Unified}} & \multicolumn{1}{c}{\textbf{Disjointed}} \\
        \cline{2-3}
         \textbf{Parsimonious} & 
            \vspace{.25cm}
            \begin{tikzpicture}[x=1cm,y=1cm]
                %   \hspace{.75cm}
                  \node[obs]                                      (X1)      {$x_1$} ; %
                  \node[obs, right=of X1]                         (X2)      {$x_2$} ; %
                  \node[latent, above=of X1, xshift=.75cm]        (Z1)      {$z_1$} ; %
                  \edge {Z1} {X1,X2} ; %
                \end{tikzpicture}  
                \vspace*{.25cm}
                & 
                \vspace{.25cm}\hspace{.75cm}\begin{tikzpicture}[x=1cm,y=1cm]
                  \node[obs]                                      (X1)      {$x_1$} ; %
                  \node[obs, right=of X1, xshift=.25cm]           (X2)      {$x_2$} ; %
                  \node[latent, above=of X1]                      (Z1)      {$z_1$} ; %
                  \node[latent, above=of X2]                      (Z2)      {$z_2$} ; %
                  \edge {Z1} {X1} ; %
                  \edge {Z2} {X2} ; %
                \end{tikzpicture}
                \vspace*{.25cm}
                \\ 
            \cline{2-3}
            \textbf{Elaborate} & 
            \vspace{.25cm}
            \begin{tikzpicture}[x=1cm,y=1cm]
                  \node[obs]                                      (X1)      {$x_1$} ; %
                  \node[obs, right=of X1]                         (X2)      {$x_2$} ; %
                  \node[latent, above=of X1, xshift=-.75cm]       (Z1)      {$z_1$} ; %
                  \node[latent, above=of X1, xshift=.75cm]        (Z2)      {$z_2$} ; %
                  \node[latent, above=of X2, xshift=.75cm]        (Z3)      {$z_3$} ; %
                  \edge {Z1} {X1,Z2} ; %
                  \edge {Z2} {X1} ; %
                  \edge {Z3} {Z2,X2} ; %
                  \edge {X2} {Z2} ; %
                \end{tikzpicture}
                \vspace{.25cm}
         & 
         \vspace{.25cm}
         \begin{tikzpicture}[x=1cm,y=1cm]
              \node[obs]                                      (X1)      {$x_1$} ; %
              \node[obs, right=of X1, xshift=.25cm]           (X2)      {$x_2$} ; %
              \node[latent, above=of X1, xshift=-.75cm]       (Z1)      {$z_1$} ; %
              \node[latent, above=of X1, xshift=.75cm]        (Z2)      {$z_2$} ; %
              \node[latent, above=of X2, xshift=0cm]       (Z3)      {$z_3$} ; %
            %   \node[latent, above=of X2, xshift=.75cm]        (Z4)      {$z_4$} ; %
              \edge {Z1} {X1,Z2} ; %
              \edge {Z2} {X1} ; %
              \edge {Z3} {X2} ; %
            %   \edge {Z4} {X2} ; %
        \end{tikzpicture} 
        \vspace{.25cm}
         \\ 
         \cline{2-3}
        \end{tabular}

        \end{center}

    %  \end{tcolorbox}
    \caption{Table 1: In these diagrams that represent the causal structure of different explanations, shaded nodes represent observables, other nodes represented postulated latent causes, and arrows represent causal relationships. Simplicity is a rich concept that likely binds together multiple intuitions simultaneously. One way of assessing simplicity is parsimony, which counts the number of parameters or latent causes invoked by an explanation. A second way of assessing simplicity is unification, which measures the degree to which a theory provides an overarching, connected account of multiple features of the world. Explanations can vary along each of these dimensions independently, so that their overall ``simplicity'' might be judged as an additive composite of the two.}
    \label{box:simplicity}
\end{figure}
     
\section{When Values become Vices}

Our approach yields a normative prescription where deviations from equal weighting lead to characteristic explanatory pathologies which canoperationalize what Ref.~\citep{cassam2016vice} calls ``vice epistemology''.

Consider the phenomenon of overgeneralization~\citep{williams2013hazards}, \emph{i.e.}, attempting to cover all examples with a single explanation rather than allowing for exceptions. This can be caused by over-valuing co-explanation relative to descriptiveness, or by over-valuing unification (since theories that have high unification will tend to have higher co-explanation when they are good fits to the data). 

What makes a weighting virtuous varies across contexts. For example, the appropriate amount of simplicity depends upon the domain \cite{huemer2009parsimony}, and there is evidence that people complexity-match~\cite{complexity_matching}, i.e., allow the perceived complexity of the explanandum to guide priors on the simplicity of the explanation. With that being said, recent empirical work has started to tie abnormal reasoning to common inferential biased that generalize across domains in a way that suggests the systemic miscalibration of values may be at fault. 

For example, as noted by Ref.~\cite{brotherton2014belief}, those prone to paranormal thinking also show susceptibility to the conjunction fallacy. This can result from overvaluing co-explanation, because labeling Linda a feminist as well as a bank teller~\citep{tversky1974judgment} provides a co-explanatory account of her political and social activities. Empirical work has also established strong individual differences in the tendency to believe conspiracies: those who believe one are more likely to believe others \citep{bruder2013measuring}. This trait is common in individuals with schizotypal disorders~\cite{darwin2011belief}, which are linked in turn to a number of other explanatory abnormalities~\cite{mclean2017association}. The finding that conspiracy-mindedness is a stable trait suggests that the its associated beliefs may be accumulated over time due, at least in part, to a systematic miscalibration in an individual's weighting of explanatory values. 

Conspiracy theories are often both abnormally co-explanatory \emph{and} descriptive \citep{vitriol2018illusion}. They account for anomalous facts which are unlikely under the ``official'' explanation (``errant data''~\cite{errant_data}; see, {\emph e.g.}, Ref.~\citep{okbomb} on Oaklahoma City bombing conspiracy theories), and show how seemingly arbitrary facts of ordinary life are correlated by hidden events~\citep{Tangherlini_2017}; Ref.~\cite{whitson2008lacking} finds that a manipulation which induces subjects to see (illusory) correlations in neutral domains like stock returns also increases beliefs in conspiracy theories. Finally, and famously, conspiracy theories are unifying: they describe a universe where everything is correlated by a network of hidden common causes---the motives and meetings of the conspirators \citep{douglas2017psychology}.
        
Valuing these features is not, in and of itself, a vice. What frequently goes wrong is the failure to balance these values against others, such as simplicity or contextual priors that urge trust in institutions and down-weight generalizations associated with racist, sexist, or antisemitic prejudice. On the surface, a conspiracy theory is quite simple; as it is unfolded, however, increasing complexity is required to explain contradictory evidence and the cover-up that has, so far, prevented it from coming to light. Such a judgement is itself open to criticism; as noted by \citep{dentith2016inferring}, some conspiracies are extremely compelling on normative grounds. Some even turn out to be true. 

This latter point gets to the heart of what makes explanation so difficult. Striking a virtuous balance between so many considerations is itself a challenging cognitive problem, one that we solve partially by social circumspection. Failures at this level might help to explain anti-vaccination movements \citep{miton2015cognitive}, COVID-19 conspiracies~\cite{shahsavari2020conspiracy}, the use of pseudoscience in extremist ideologies \cite{o2018seduction}, and science denialism \citep{rutjens2018attitudes}. While these beliefs are in part formed and maintained by social processes in addition to epistemic ones, their core logic often appeals to many of the same explanatory imbalances as conspiracy theories do. These interact with individual-level predispositions, including what are usually taken to be pathologies of thought. One avenue for future research is how social processes may serve to maintain, accentuate, or exploit individual-level explanatory imbalances.\footnote{Vices of overvaluation naturally co-exist with other biases, such as, for example, omission bias in the case of anti-vaccine movements~\citep{miton2015cognitive}.} 

% These forces interact with individual-level predispositions, including what are usually taken to be pathologies of thought. For example, the preference for conspiracy-mindedness seen in individuals with schitzotypal disorders~\cite{darwin2011belief} may help us understand the neurological mechanisms in play in the perception of different explanatory values. 

 \begin{figure}
    \centering
    \begin{tcolorbox}
        \textbf{Box 3: Three Stages of Explanation} \vspace{3mm} \newline
            \begin{tabularx}{\textwidth}{
                >{\raggedleft\arraybackslash}X
                >{\centering\arraybackslash}X
                >{\centering\arraybackslash}X
                >{\centering\arraybackslash}X }
                & \textbf{1. Generation} &\textbf{2. Selection} & \textbf{3. Evaluation}  \\ [0.5ex] 
                \cmidrule(lr){2-2}\cmidrule(lr){3-3}\cmidrule(lr){4-4} 
                \multicolumn{1}{r}{\textbf{Observations:}} & fixed & fixed & variable \\ [1ex]
                \multicolumn{1}{r}{\textbf{Explanations:}} & variable & fixed & fixed \\ [1ex]
                \end{tabularx} 
        
            Explanation decomposes into three stages: we generate explanations (or receive them from others), select among them, and re-evaluate them based on subsequent experience. Our piece has focused on how values influence selection, but they are equally important in generation and re-evaluation. \\
            
            Values help us decide which experiences to seek out next. Co-explanation may lead me to look in places where I expect the data to be correlated under a favored hypothesis. Descriptiveness, conversely, may lead one to look for key counter-examples, in the style of Karl Popper's falsification \citep{popper2005logic}. \\
            
            Values also act at the generation stage. Ref.~\cite{gelman2013philosophy} describes scientific hypothesis formation as descriptiveness-driven, where explanations are updated in response to outliers: one produces an explanation, looks for places that it fails, and tries to update it in response. Co-explanation in the generation phase can also look like Peircean abudction~\cite{mcauliffe2015did}. The cycle, augmented to include the gathering of data, is used by both children \cite{bonawitz2012occam} and adults \citep{bramley2017formalizing}. \\
            
            A descriptiveness-driven cycle need not be virtuous: updating an explanation may increase its descriptiveness at the cost of theoretical values such as unification or simplicity. Kuhn's ``paradigm shift'' is driven in part by the decreasing simplicity of the standard paradigm: as anomalies arrive, more and more epicycles are required to explain them~\cite{kuhn2012structure}. \\
            
            Rather than looking for outliers because we value descriptiveness, one may look for correlations because we value co-explanation. A person who subscribes to a ``unifying'' group stereotype, for example, may ask if people belong to the group in question when they show its characteristic behaviors. More virtuously, co-explanation can drive scientists to compare evidence across different domains. \\
            
            Empirical values also direct attention at a more basic, cognitive level. Ref~\cite{itti2009bayesian} find that descriptiveness draws the eye to outliers: attention to a group of pixels correlates with deviations from its predicted distribution. Co-explanation, conversely, draws attention to patterns that constitute gestalts \cite{desolneux2007gestalt}. \\
            
            Theoretical values, conversely, are crucial to how we go about generation, because we cannot consider every  explanation. Parsimony and contextual considerations help us reject certain types of explanations out of hand~\cite{gopnik2017changes}, while unification makes the world itself easier to remember and describe. Theoretical values can sometimes be a vice, and we often fail to generate good explanations even when we have the ability to recognize them \cite{rothe2018people}.
     \end{tcolorbox}
     \caption{}
     \label{box:dynamics}
 \end{figure} 

\section{Concluding remarks}

% shenhav2017toward removed because we needed to add citations requested by a referee, and we're over limit
Framing explanatory values as components of a Bayesian inference is a form of  \emph{rational analysis}, which seeks to understand mental states in terms of the computational goals they help agents achieve \citep{lieder2019resource,chater1999ten}. Such an approach has been applied to a wide range of subjective states such as representativeness \citep{tenenbaum2001rational}, suspicious coincidence \citep{griffiths2007mere}, randomness \citep{griffiths2018subjective}, tip of the tongue \citep{brown1991review}, boredom and flow \citep{chater2019boredom}, mental effort \citep{kurzban2013opportunity}, and curiosity \citep{loewenstein1994psychology,pathak2017curiosity,loewenstein2020curiosity}. Of these, explanation is most closely related to curiosity. If curiosity drives us to seek answers to salient questions \citep{golman2015curiosity} and to make sense of the world around us \cite{chater2016under}, then explanatory values are the subjective states that signal, often in compelling hedonic form \citep{gopnik1998explanation}, when good answers have been found.

A Bayesian framing naturally centers on an explanation's ability to predict observed data. Explanation is more than prediction, however, and other features are necessary to satisfy the many social, cognitive, and practical constraints that bear on the practice. A highly-predictive black box, for example, is not something that we can evaluate in terms of theoretical constructions such as parsimony or unification. Even when the black box is opened, what is inside may be so far from ``virtuous'' in the human sense that it scarcely counts as an explanation at all---even if it is intelligible in a literal sense. There is something more basic yet, of course: an explanation must be intelligible before we can ask about its value. This is part of the ``explainability crisis'' in machine learning~\cite{pmlr-v81-selbst18a} and is crucial to understanding, and thereby closing, the gap between human and artificial intelligence~\cite{ai_tests,mao2019neuro}. While a rational analysis of explanatory values is an important first step, further work is needed to address the intelligibility problem.

All explanation occurs against a background of folk theories, world-views, and  explanations that have come before \cite{kitcher1989explanatory}. This opinion suggests that it may be possible to enumerate ``atomic'' explanatory values, and that the history of explanation is largely the history of their relative emphasis. Given that explanations emerge in a social context, however, we might also expect new values---especially theoretical ones---to appear over time. This dynamical, contextual natural of explanation are clarifies why explanations seem more valuable when they co-explain phenomena that, on the basis of previous understanding, were conceptually distant (see Box 3). Indeed, conceptual distance and co-explanation may be two sides of the same coin: what is conceptually distant may just be what, with our current explanations, we can not co-explain. 
% Explanation is dynamic, and any explanation is evaluated in a context of prior explanations. 
% % https://english.stackexchange.com/questions/140639/dynamical-vs-dynamic
% Among other things, this clarifies why explanations seem more valuable when they co-explain phenomena that, on the basis of previous understanding, were conceptually distant. Indeed, conceptual distance and co-explanation may be two sides of the same coin: what is conceptually distant is just what, with our current explanations, we can not co-explain. 
Another important aspect of the dynamical side of explanation is the role of prediction out of sample, \emph{i.e.}, re-evaluation when new information arrives. Because predicting the future is much more challenging than accounting for what is already known, doing so can be a powerful source of (empirical) value.

The importance of conceptual distance and the power of confirmation by unexpected data come together in ``consilience''. Consilience is an explanatory value introduced by \cite{whewell1858history} in the 19th Century to describe features of scientific explanations that, he argues, both are, and ought to be, prized by the community. ``The Consilience of Inductions'', Whewell writes, ``takes place when an induction, obtained from one class of facts, coincides with an induction, obtained from another class... Such a coincidence of untried facts with speculative assertions cannot be the work of chance, but implies some portion of truth in the principles on which the reasoning is founded.'' Indeed, for Whewell, consilience carries simplicity and unification along for the ride: for consilient explanations, ``all the additional suppositions tend to simplicity and harmony... the system becomes more coherent as it is further extended. The elements which we require for explaining a new class of facts are already contained in our system. Different members of the theory run together, and we have thus a constant convergence to unity.''

Whewell is far from the only writer to draw attention to the general features of what makes for good explanation, and a significant part of social interaction involves debating and arguing for different explanations on the basis of the values they exhibit~\cite{mercier2017enigma}. The implicit bargain for this paper is that such values may be amenable to an analysis in terms of basic, atomic units active in similar ways across a great variety of domains. Once identified, these units can provide a new way to understand how people make sense of the world.

\section*{Acknowledgements}

We acknowledge the support of the John Templeton Foundation, and Jaan Tallinn via the Survival and Flourishing Fund.

% \vspace{2cm}
% \textbf{requirements:} \url{https://www.cell.com/trends/cognitive-sciences/authors}

\newpage 

\input{outstanding}

\clearpage
\pagebreak \newpage
% \singlespacing
\onehalfspacing
% \setlength\bibsep{0pt}
% \section*{References} \label{sec:references}
% \renewcommand{\section}[2]{}%
% \nocite{*}
%\bibliographystyle{unsrt}
%\bibliography{main}

\end{document}

%% file: highlights.tex
\begin{figure}
    \begin{tcolorbox}
    \begin{center}
        \textbf{Highlights}
    \end{center}
    \begin{itemize}
        \item Recent experiments show that we value explanations for many reasons, such as predictive power and simplicity.
        \item Bayesian rational analysis provides a functional account of these values, along with concrete definitions that allow us to measure and compare them across a variety of contexts including visual perception, politics, and science.
        \item These values include descriptiveness, co-explanation, unification, power, and simplicity, and fall into two groups: the first two are associated with the evaluation of explanations in the light of experience, while the latter concern the intrinsic features of an explanation.
        \item Failures to explain well can be understood as imbalances in these values: a conspiracy theorist, for example, may over-rate co-explanation relative to simplicity, and many similar ``failures to explain'' that we see in social life may be analyzable at this level.
    \end{itemize}
    \end{tcolorbox}
\end{figure}

%% file: outstanding.tex
\begin{figure}
    \begin{tcolorbox}
    \begin{center}
        \textbf{Outstanding Questions}
    \end{center}
    \begin{itemize}
    \item How are the atomic elements of explanatory values perceived and combined by the mind? Why these elements, and not others? Where do they come from in evolutionary and cultural time?
    \item To what extent are values determined during early childhood development, versus learned in later life? Can people change their values in response to experience or teaching?
    \item How do explanatory values influence the cultural evolution of explanations?
    \item What determines the categories (\emph{i.e.} variables) over which explanatory values are evaluated? How do these co-evolve with explanations? Are these categories determined by other forces, or are they---at least partially---determined by explanatory considerations themselves?
    \item How universal are these values? How much of the difference between individual preferences for explanations is driven by domain-general explanatory values versus contextual priors?
    \item What is the connection between organic brain diseases and imbalanced explanatory values? What can this tell us about how the neurological basis of these values and the manner in which they are assessed?
    \item To what extent are social movements associated with pathological beliefs (such as extremist ideologies) driven by explanatory imbalance? Does participation in such a movement reinforce such imbalances?
    \item To what extent are values simply a means of achieving the practical goal of prediction? What other roles do they play in human life?
    \item What is the relationship between moral and practical explanation?
    \item What makes an explanation intelligible ``to us''? 
    \item How can we enable machines to explain as well as predict? Can explanatory values help the task of bridging the gap between human and artificial intelligence? 
    \end{itemize}
    \end{tcolorbox}
\end{figure}